\documentclass{article}
\usepackage{fullpage}
\usepackage{graphics,psfrag,epsfig}
\usepackage{amsfonts,latexsym,eucal,amsmath,amsthm,amssymb}

\begin{document}

\newcommand{\rum}{\rule{0.5pt}{0pt}}
\newcommand{\rub}{\rule{1pt}{0pt}}
\newcommand{\rim}{\rule{0.3pt}{0pt}}
\newcommand{\numtimes}{\mbox{\raisebox{1.5pt}{${\scriptscriptstyle \times}$}}}
\newcommand{\optprog}[2]
{%
  \noindent\mbox{}\\[0cm]
  \noindent\fbox{%
  \begin{minipage}{0.955\linewidth}
    \mbox{}\\[-0.5cm]
    #1\\[#2]
  \end{minipage}
  }
  \noindent\mbox{}\\[-0.2cm]
}

\renewcommand{\refname}{References}

\twocolumn[%
\begin{center}
{\Large\bf 2-State 3-Symbol Universal Turing Machines Do Not Exist \rule{0pt}{13pt}}\par
\bigskip
Craig Alan Feinstein \\ {\small\it 2712 Willow Glen Drive, Baltimore, Maryland 21209\rule{0pt}{13pt}}\\
\raisebox{-1pt}{\footnotesize E-mail: cafeinst@msn.com, BS"D}\par
\bigskip
{\small\parbox{11cm}{%
\bigskip \noindent \textbf{Abstract:} In this brief note, we give a simple
information-theoretic proof that 2-state 3-symbol universal Turing machines cannot possibly exist, unless
one loosens the definition of ``universal".

\bigskip \noindent \textbf{Disclaimer:} This article was authored by Craig
Alan Feinstein in his private capacity. No official support or endorsement by the U.S. Government is
intended or should be inferred.\rule[0pt]{0pt}{0pt}}}
\bigskip
\end{center}]{%

In May of 2007, Wolfram Research offered a prize to anyone who could answer the question of whether
a particular 2-state 3-symbol Turing machine is universal. In October of 2007, Wolfram Research announced
that Alex Smith, a student at the University of Birmingham, proved that the particular 2-state 3-symbol
Turing machine is universal \cite{b:Wolpr}. But not every expert in the field of theoretical computer
science was convinced that Alex Smith's proof was valid \cite{b:FOM}. In this note, we give a simple
information-theoretic proof that 2-state 3-symbol universal Turing machines cannot possibly exist,
unless one loosens the definition of ``universal":

A universal Turing machine must be able to perform binary operations like OR, AND, XOR, etc., between bits,
and its tape-head must have the freedom to move left or right independent of the binary operations, in order
to simulate other Turing machines with this property. This 
implies that the tape-head of a universal Turing machine must be able to keep track of at least 
three bits of information at a time, at least two for binary operations and at least one for the 
direction that the tape-head moves. 

The tape-head of a 2-state 3-symbol 
Turing machine can only keep track of $\log_2 (2 \times 3)$ bits of information at a time, which is less 
than three bits of information; therefore, no 2-state 3-symbol universal Turing machine can possibly exist,
unless one loosens the definition of ``universal".

}

\end{document}